\documentstyle[aps,prl,epsf,twocolumn]{revtex}

\newcommand{\vc}[1]{{\mbox{\boldmath $ #1$}}} 

\begin{document}
\title{Statistical anisotropy of magnetohydrodynamic turbulence}
\author{Wolf-Christian M\"uller, Dieter Biskamp}
\address{Centre for Interdisciplinary Plasma Science,\\
Max-Planck-Institut f\"ur Plasmaphysik, Euratom Assoziation, 85748 Garching, Germany}
\author{Roland Grappin}
\address{Observatoire de Paris-Meudon, LUTH, 92195 Meudon, France}
\maketitle
\begin{abstract}
Direct numerical simulations of decaying and forced magnetohydrodynamic (MHD) turbulence without and with mean magnetic field are analyzed
by higher-order two-point statistics. The turbulence exhibits statistical anisotropy with respect to the direction of the local magnetic field 
even in the case of global isotropy.
A mean magnetic field reduces the parallel-field dynamics while in the perpendicular 
direction a gradual transition towards two-dimensional MHD turbulence is observed with $k^{-3/2}$ inertial-range scaling of the perpendicular 
energy spectrum. 
An intermittency model based on the Log-Poisson approach, $\zeta_p=p/g^2 +1 -\left(1/g\right)^{p/g}$, is able   
to describe the observed structure function scalings. 
\end{abstract}
\draft
Turbulence is the natural state of many plasma flows observed throughout the universe, its
statistical properties being essential for the theoretical understanding
of, e.g., star-forming regions in the interstellar medium, the convection in planetary and stellar interiors, 
as well as the dynamics of stellar winds. The solar wind, in particular,  represents the only source of in-situ measurements, 
since laboratory experiments are far from generating fully-developed turbulence at high magnetic Reynolds numbers.
A simplified nonlinear model of turbulent plasmas is incompressible magnetohydrodynamics (MHD) \cite{biskamp:book}.
In this framework the kinetic nature of microscopic processes responsible for, e.g., energy dissipation, is neglected
when studying the fluid-like macroscopic plasma motions.

The spatial similarity of incompressible MHD turbulence is usually investigated by considering
two-point statistics of the Els\"asser variables $\vc{z}^\pm=\vc{v}\pm\vc{B}$ \cite{elsaesser:mhd} 
combining velocity $\vc{v}$ and magnetic field $\vc{B}$ (given in Alfv\'en-speed units).
Restricting consideration to turbulence with small cross helicity $H^{\mathrm{C}}=\int_V\mathrm{d}V(\vc{v}\cdot\vc{B})$, 
$V$ being the volume of the system, allows to set $\vc{z^+\simeq z^-=z}$.
With $\delta z_\ell=\left[\vc{z}(\vc{r}+\vc{\ell})-\vc{z}(\vc{r}) \right]\cdot \vc{\ell}/\ell$ the
longitudinal isotropic structure functions of order $p$ are defined as $S_p(\ell)=\langle\delta z_\ell^p \rangle$,
the angular brackets denoting spatial averaging. 
The structure functions exhibit self-similar scaling $S_p(\ell)\sim\ell^{\zeta_p}$ in the inertial range where the dynamical influence of dissipation, 
turbulence driving and system boundaries is weak.

The inertial range has been introduced in Kolmogorov's K41 phenomenology of incompressible hydrodynamic turbulence 
\cite{kolmogorov:k41a,kolmogorov:k41c} which assumes a spectral energy-cascade driven by the break-up of turbulent eddies. 
This leads to the experimentally well-verified energy-spectrum $E(k)\sim k^{-5/3}$ \cite{frisch:book} corresponding to $\zeta_2=2/3$. 
Iroshnikov and Kraichnan (IK) \cite{iroshnikov:ikmodel,kraichnan:ikmodel} included the effect of 
a magnetic field by founding the energy-cascade on the mutual scattering of Alfv\'en waves triggered by velocity fluctuations. 
The IK picture phenomenologically yields $E(k)\sim k^{-3/2}$, i.e., $\zeta_2=1/2$.
 
The validity of the two phenomenologies in MHD turbulence is still under discussion. Two-dimensional direct numerical simulations (DNS)  
support the IK picture \cite{biskamp_schwarz:2dmhd,pol_pouq:2dmhd_structurefunc} 
while three-dimensional simulations exhibit K41 scaling behavior \cite{biskamp_mueller:3dmhdscalprop}.
Analytical results \cite{pol_pouq:4/3lawII} also suggest $\zeta_3=1$, consistent with 
K41 energy-spectra measured in the solar wind \cite{leamon_etal:solwindspec}.
A recent phenomenology of Goldreich and Sridhar \cite{goldreich_sridhar:gs2} 
postulates a balance between K41 and IK energy
cascades and accounts for the local anisotropy induced by $\vc{B}$. However, DNS which claim to support this picture 
\cite{maron_goldreich:anisomhd,cho_lazarian_vishniac:anisomhd} suffer from moderate numerical resolution, making the 
identification of self-similar scaling ranges difficult.

In this Letter we examine three-dimensional pseudospectral DNS of decaying isotropic MHD turbulence and of
driven turbulent systems with mean magnetic field $\vc{B}_0$ at comparably high resolutions of up to $512^3$ collocation points. 
The structure functions are found to be anisotropic with respect to the local magnetic field. The effect increases with magnetic-field
strength, reducing the spatial intermittency of the turbulence in the parallel-field direction while rendering the system quasi-two-dimensional  
perpendicular to $\vc{B}$. An intermittency model based on the Log-Poisson approach agrees well with the observed 
structure-function scalings.

The simulations are performed by numerically solving the incompressible MHD equations
\begin{equation}
\partial_t\vc{\omega} -\nabla\times(\vc{v}\times\vc{\omega}+ \vc{j}\times\vc{B})=\mu_{\nu}(-1)^{\nu-1}\Delta^{\nu}\vc{\omega},
\label{2}
\end{equation}
\begin{equation}
\partial_t\vc{B} -\nabla\times(\vc{v}\times\vc{B})=\eta_{\nu}(-1)^{\nu-1}\Delta^{\nu}\vc{B},\label{1}
\end{equation}
\[
\vc{\omega} = \nabla \times \vc{v}, \quad \vc{j}=\nabla\times\vc{B},\quad \nabla\cdot\vc{v}=\nabla\cdot\vc{B}=0,
\]
in a periodic cube with a pseudo-spectral method using spherical mode-truncation to reduce aliasing effects \cite{vincent_meneguzzi:simul}. 
All simulations comprise about $9$ eddy-turnover times, a turnover time being defined as the period required to reach maximal energy
dissipation when starting from smooth fields. 
The initial conditions in the decaying case are characterized by
global equipartition of kinetic and magnetic energy $E^{\mathrm{K}}=E^{\mathrm{M}}=0.5$ with a spectral distribution $\sim \exp(-k^2/k_0^2)$, $k_0=4$, and 
random phases. 
The driven runs start from a quasi-stationary state 
with $E^{\mathrm{K}}\simeq 0.75$, $E^{\mathrm{M}}\simeq 0.8$ generated by forcing the system for $24$ turnover times. 
The forcing is realized by freezing all modes in a sphere of radius $k_{\mathrm{f}}=2$, allowing energy transfer to higher wavenumbers by nonlinear 
interactions only.       
The magnetic helicity in all simulations is finite, 
$H^{\mathrm{M}}={\int_V\mathrm{d}V(\vc{A}\cdot\vc{B})\approx 0.7} H^{\mathrm{M}}_{\mathrm{max}}$ with the magnetic vector potential
$\vc{A}$ and $H^{\mathrm{M}}_{\mathrm{max}}\sim E^{\mathrm{M}}/k_0$. 
For the driven cases the $\vc{B}_0$-component which renders $H^{\mathrm{M}}$ gauge-dependent has been subtracted and $k_0\simeq k_{\mathrm{f}}$. 
The cross helicity $H^{\mathrm{C}}$ is approximately zero in the decaying system and $\simeq 0.12$ for the driven runs, mildly fluctuating 
around the respective value.
The generalized magnetic Prandtl number $\mu_{\nu}/\eta_{\nu}$ is equal to unity with hyperdiffusive dissipation operators 
$\nu=2$, $\mu_2=\eta_2=3\times10^{-8}$. 
Test runs with $\nu=1$ show no notable difference in the reported results.   

Weak-turbulence theory \cite{galtier_nazarenko_pouquet:weakMHDturb} and numerical simulations 
\cite{shebalin_matthaeus:aniso,grappin:aniso,ng_bhattacharjee:gsaniso} show that 
the IK cascade is spatially anisotropic, the underlying three-wave interactions distributing energy predominantly perpendicular to $\vc{B}$.
The associated spectral dependence on a mean magnetic field has been studied numerically as well as in the reduced-MHD framework 
(\cite{oughton_matthaeus:anisormhd} and references therein) while
statistical anisotropy has been found in the second-order structure functions of MHD turbulence \cite{milano_matthaeus:anisomhd}.
In order to assess this anisotropy by higher-order statistics, parallel and perpendicular structure functions are 
calculated with field increments, $\delta z_\ell$, taken in the corresponding directions with respect to the local  
magnetic field. This is in contrast to isotropic structure functions where the $\delta z_\ell$ are measured without preferred direction. 
The local magnetic field on the increment length-scale, defining field-parallel and field-perpendicular directions at each spatial co-ordinate,
is found by applying a top-hat filter of width $\ell$, i.e., by multiplying $\vc{B}$ with the 
Fourier-space filter kernel $G_\ell(k)=\sin(k\ell/2)/(k\ell/2)$. 
The obtained structure functions, computed using $|\delta z_\ell|$ to avoid cancelation effects,
exhibit inertial-range scaling with exponents $\zeta_p$ which 
can be determined up to order $p=8$ by a saddle-point in the logarithmic derivative. The results have been cross-checked using 
the extended self-similarity property \cite{benzi:ess} of the data. 

\begin{figure}
\epsfxsize=3truein %13truecm
\epsfbox{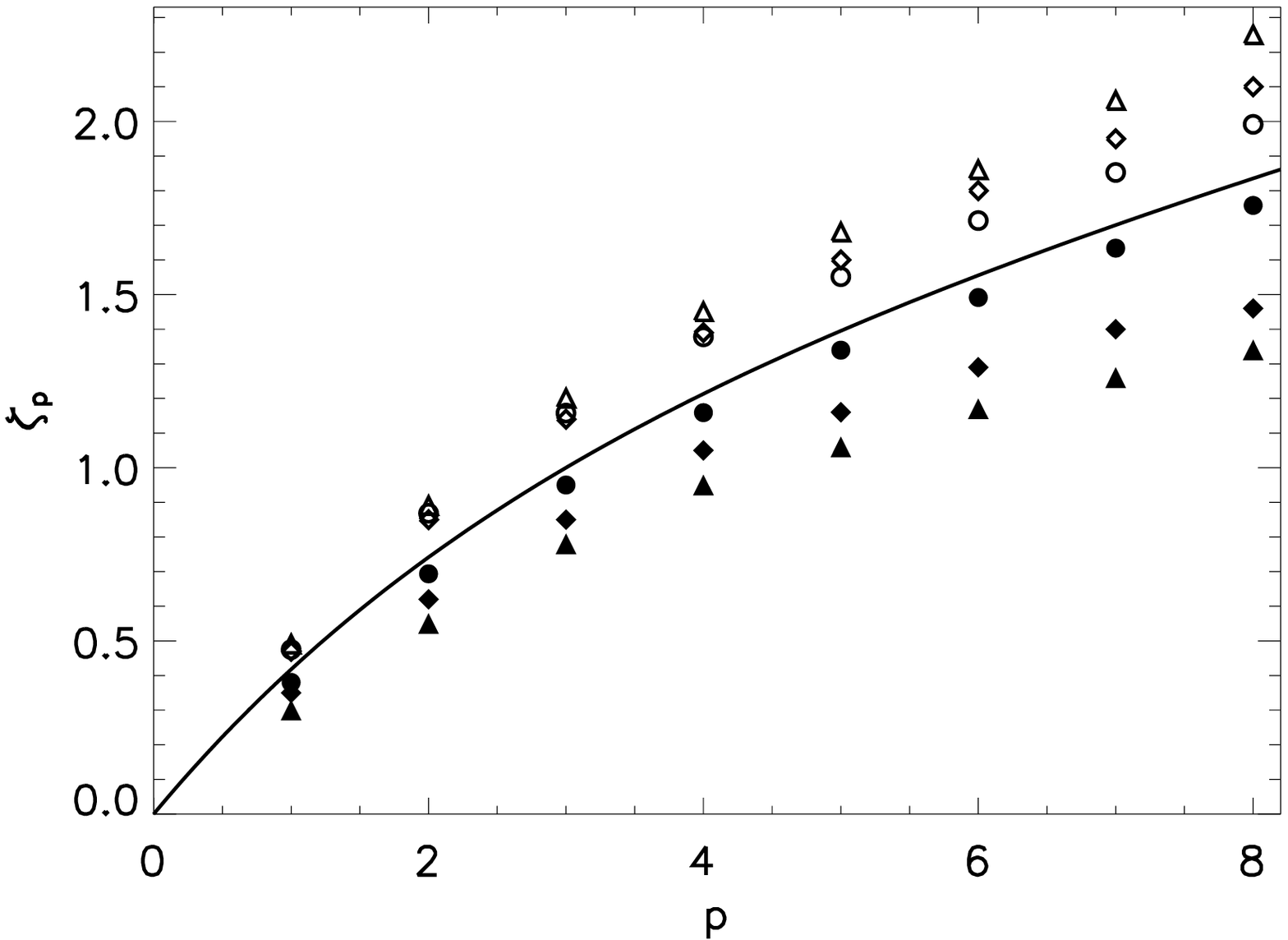}
\caption[Scaling exponents $\zeta_p$ of perpendicular (filled symbols) and parallel (open symbols) structure-functions 
$S_p(\ell)=\langle |\delta z_\ell|^p\rangle$ for $B_0=0,5,10$ (resp. circles, diamonds, triangles) together with
isotropic scalings from 3D-DNS 
(solid line, \protect\cite{mueller_biskamp:3dmhdscale}). Error-bars are given in Fig. \ref{f3}.]{}  
\label{f1}
\end{figure}

Fig. \ref{f1} shows that the decaying system with $B_0=0$ ($512^3$ collocation points) is statistically anisotropic.
The field-perpendicular scalings display increased 
intermittency compared to the isotropic structure functions. Assuming that the formation of intermittent structures 
primarily depends on the turbulent energy available at the corresponding spatial scales, the observation is consistent  
with both the IK and the K41 cascade being strongest perpendicular to $\vc{B}$. 
The former due to the inherent anisotropy of Alfv\'en-wave scattering, the latter because field lines resist bending but can be shuffled 
easily by eddies perpendicular to the local field. Consequently, the field-parallel energy transfer is depleted leading to less intermittent parallel 
scalings.

\begin{figure}
\epsfxsize=3truein %13truecm
\epsfbox{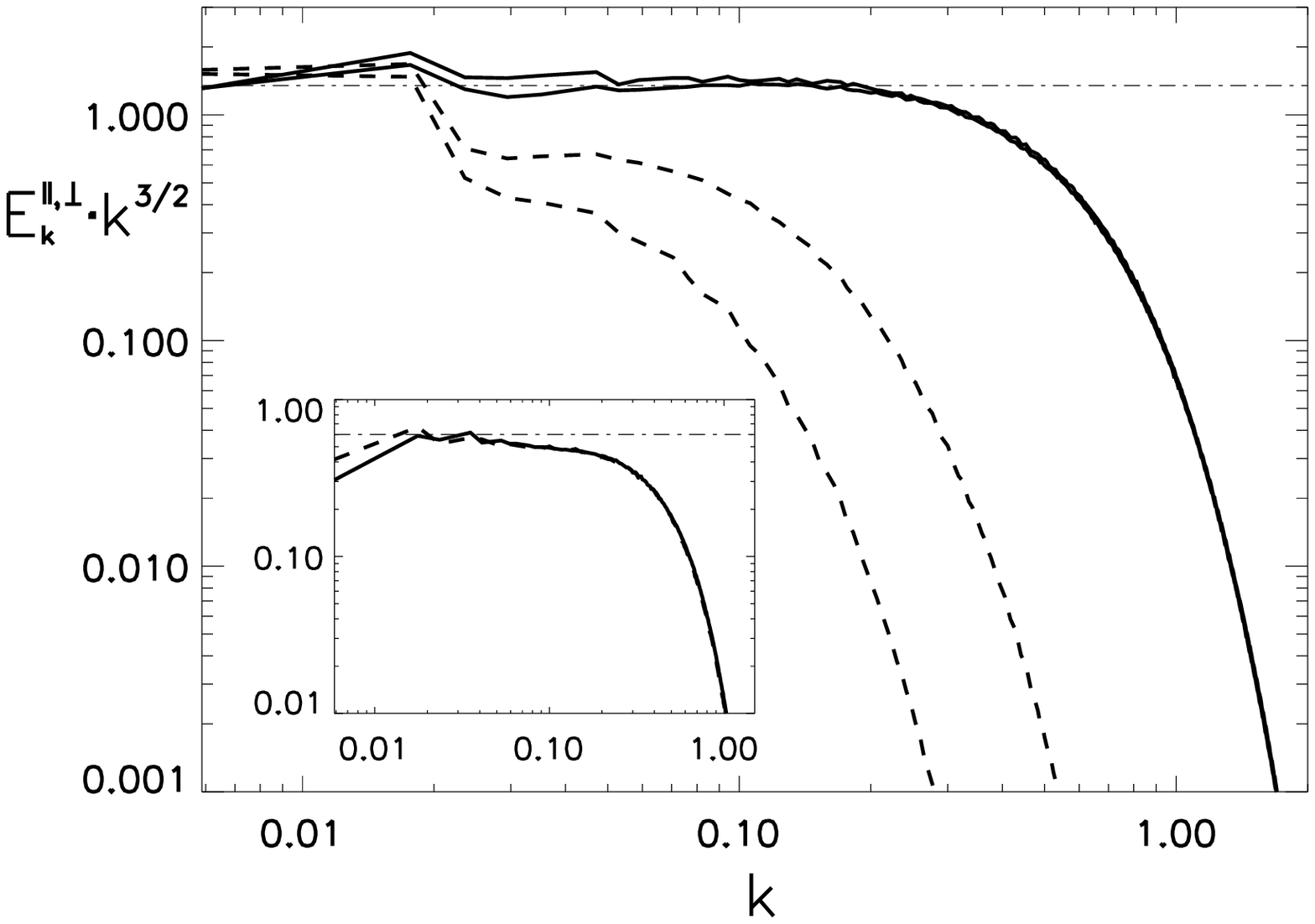}
\caption[Normalized, time-averaged parallel (dashed) and perpendicular (solid) energy spectra compensated with $k^{3/2}$ 
for $B_0=0$ (inset), $B_0=5$ (lower solid line, upper dashed line) and $B_0=10$.]{}  
\label{f2}
\end{figure}

The dependence of statistical anisotropy on the magnetic-field strength is examined in forced-turbulence simulations with mean magnetic 
field ($B_0=5$,$10$) allowing a reduction of numerical resolution in the mean-field direction to respectively $256$ and $128$ collocation points
(cf. Fig. \ref{f2}). 
The quasi-stationary forced systems assume constant energy ratios of mean field to turbulent fluctuations. 
The parallel scalings shown in Fig. \ref{f1} display decreasing intermittency with raising 
$B_0$, i.e., an asymptotic approach towards a straight line. Referring to Kolmogorov's refined-similarity hypotheses \cite{kolmogorov:refined}, this 
is equivalent to higher parallel-field homogeneity of small-scale dissipative structures, i.e., current and vorticity micro-sheets, 
due to their stronger alignment along $\vc{B}_0$.  
The perpendicular statistics (Fig. \ref{f1}) become increasingly two-dimensional, getting close to values found in
DNS of two-dimensional MHD turbulence (The exponents for $B_0=10$ and those obtained from 2D-DNS \cite{pol_pouq:2dmhd_structurefunc}
coincide within the error-margin.). 
The asymptotic state results from micro-sheet alignment along $\vc{B}_0$ which decreases the spatial extent of the
sheets in the field-perpendicular direction to quasi-one-dimensional dissipative ribbons. 

The $\zeta_2$-exponents  are related to the inertial-range scaling of the one-dimensional energy spectra 
$E_k=1/2\int\mathrm{d}k_1\int\mathrm{d}k_2(|\vc{v}_{\vc{\scriptstyle k}}|^2+|\vc{B}_{\vc{\scriptstyle k}}|^2)$ 
with $k_1$, $k_2$ spanning planes perpendicular to the component of $\vc{k}$ associated with the spatial increment $\vc{\ell}$. 
For the field-parallel and field-perpendicular spectra this gives $E_k^{\parallel ,\perp}\sim k^{-(1+\zeta_2^{\parallel ,\perp})}$. 
Fig. \ref{f1} yields $E_k^{\parallel}$-exponents in the range $[-1.8,-1.9]$ while the $E_k^{\perp}$-scaling  changes from 
K41, $\sim k^{-5/3}$, to IK, $\sim k^{-3/2}$, with increasing $B_0$. 
This agrees with DNS of two-dimensional MHD turbulence and suggests that, contrary to the three-dimensional case where K41 scaling is observed, 
the restriction to a quasi-two-dimensional geometry increases the importance of the inherently two-dimensional Alfv\'en-wave cascade (IK) compared to 
the eddy-breakup process (K41). 

In Fig. \ref{f2}, $E_k^{\parallel}$ and $E_k^{\perp}$ with respect to the fixed $\vc{B}_0$-axis are given for $B_0=0,5,10$.
The spectra are time-averaged over $4$ eddy turnover-times and normalized in amplitude assuming a K41 dependence on the mean 
energy-dissipation $\varepsilon=-\dot E$ as $\sim\varepsilon^{2/3}$. Wavenumbers are normalized with the generalized K41 dissipation-length 
$\ell_{\mathrm{D}}=(\mu^3/\varepsilon)^{1/(6\nu-2)}$. The normalization, though unnecessary for the driven runs, allows comparison with 
the decaying case shown in the inset in Fig. \ref{f2}. 
For $B_0=0$ the parallel and perpendicular energy spectra differ slightly at largest scales where the few 
involved Fourier modes do not isotropize perfectly. The inertial range exhibits K41 scaling which leads to a clear deviation from the horizontal
under the applied $k^{3/2}$ compensation.
For finite $B_0$ the spectra display a marked anisotropy in agreement with the perpendicular and parallel structure functions. 
With growing $B_0$ the $E_k^{\perp}$ asymptotically follow IK scaling while the $E_k^{\parallel}$ indicate an increasing depletion of small-scale 
turbulence.
The field-parallel damping results from the stiffness of magnetic field 
lines in agreement with the picture of field-aligned dissipative structures. This corresponds to 
an increase of the associated dissipation length \cite{grappin_mangeney:anisosolarwind}.  
The different amplitudes of $E_k^{\parallel}$ and $E_k^{\perp}$ beyond the forcing range $k\gtrsim 0.02$ have been found similarly in
shell-model calculations of anisotropic MHD turbulence \cite{carbone_veltri:anisomhdshell} 
resulting from an equilibrium between field-perpendicular and 
isotropic energy cascades. 

Intermittency, the departure of turbulence from strict spatial self-similarity, leads to   
`anomalous' non-linear behavior of the $\zeta_p$. The Log-Poisson intermittency model \cite{she_leveque:model}
reproduces these experimental and numerical findings in hydrodynamic and MHD turbulence very well. Its generic form 
$\zeta_p=(1-x)p/g+C_0(1-\left[1-x/C_0 \right]^{p/g})$ \cite{grauer_krug:mhdsl,pol_pouq:mhdsl} depends on
the codimension $C_0$ of the most singular dissipative structures (in three-dimensions $C_0=2$ for filaments, $C_0=1$ for micro-sheets), 
the scale dependence of dissipation in these structures $\varepsilon_\ell^{(\infty)}\sim\ell^{-x}$ 
and the phenomenological non-intermittent scaling $\delta z_\ell\sim\ell^{1/g}$ ($g=3$ for K41, $g=4$ for IK).
Usually, $x$ and $g$ are linked by assuming equal scaling of the timescale $t_\ell^\infty$ of 
$\varepsilon_\ell^{(\infty)}\sim E^\infty/t_\ell^\infty$ and the nonlinear transfer time $t_\ell^{\mathrm{NL}}$ of the energy cascade 
$\varepsilon \sim \delta z_\ell^2/t_\ell^{\mathrm{NL}}$ yielding $x=2/g$. 
Here, $E^\infty$ denotes the amount of energy dissipated in the most singular structures.
The successful hydrodynamic She-L\'ev\^eque formula \cite{she_leveque:model} 
results from $C_0=2$, $g=3$ while isotropic structure-function scalings
in DNS of three-dimensional MHD turbulence are well reproduced with $C_0=1$, $g=3$ \cite{mueller_biskamp:3dmhdscale}.

\begin{figure}
\epsfxsize=3truein %13truecm
\epsfbox{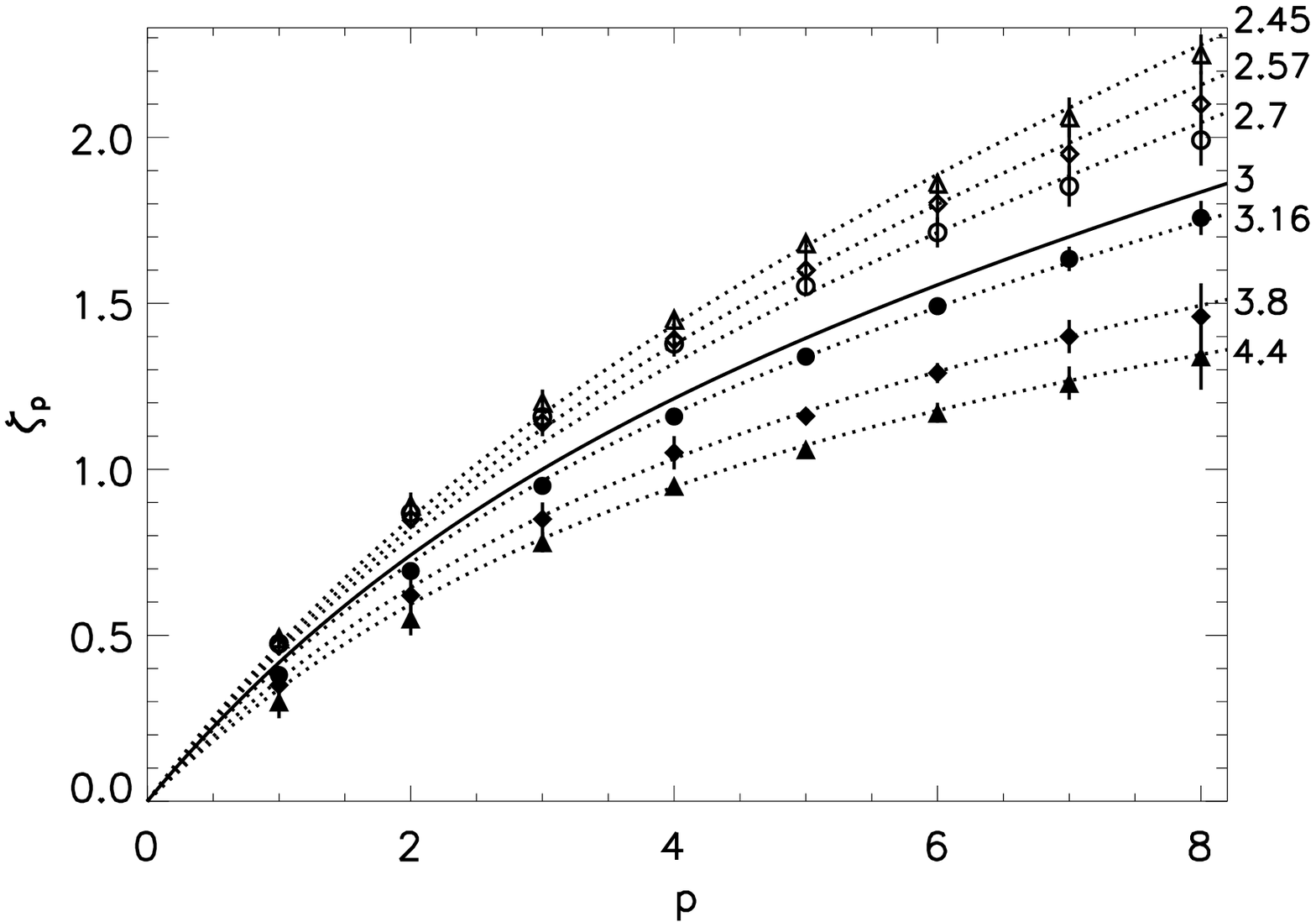}
\caption[Scaling results as in Fig. \ref{f1} combined with predictions of Eq. (\ref{anmodel}) (dotted lines). The numerical values of 
$g$ are given next to the respective curves, $g=3$ corresponds to the isotropic MHD intermittency model \protect\cite{mueller_biskamp:3dmhdscale}.]{}  
\label{f3}
\end{figure}

To model statistical anisotropy we extend the approach presented in \cite{mueller_biskamp:3dmhdscale} by 
dropping the plausible but not mandatory scaling-equality of $t_\ell^{\mathrm{NL}}$ and $t_\ell^{\infty}$. 
Instead, $t_\ell^\infty$ is fixed to the K41 timescale, $t_\ell^{\infty}\sim \ell/\delta z_\ell\sim\ell^{1-1/g}$, which together with $C_0=1$ leads to  
\begin{equation}\zeta_p=p/g^2 +1 -\left(1/g\right)^{p/g} \label{anmodel}\,.\end{equation}
Fig. \ref{f3} shows the predictions of Eq. (\ref{anmodel}) with the corresponding numerical values of $g$. 
The isotropic MHD intermittency model based on K41 scaling \cite{mueller_biskamp:3dmhdscale} is denoted by the solid line in 
Figs. \ref{f1} and \ref{f3}. 
For increasing $B_0$ the limiting value of $g$ in the parallel direction is $1$, standing for  
spatially homogeneous dissipation. The asymptotic perpendicular exponents should be reproduced by $g=4$ to be consistent 
with the IK scaling observed in this work and in DNS of two-dimensional MHD turbulence.
The fact that the observed perpendicular exponents for $B_0=10$ correspond to the model with $g\approx 4.4$  
can be ascribed to the simplicity of the approach which nevertheless captures the basic physics of the system.

By detaching  $t_\ell^{\mathrm{NL}}$ and $t_\ell^{\infty}$ the strengths of 
field-perpendicular and parallel cascades can be modified without affecting the mechanism of most singular dissipation.
The quantity  $g/3$ expresses the cascade strength relative to the isotropic K41 case as can be seen by writing 
a modified K41 transfer-time 
$t_\ell^{\mathrm{NL}}\sim \left(\ell/\ell_0\right)^\chi\left(\ell/\delta z_\ell\right)$ introducing an arbitrary reference length $\ell_0$ and the 
dimensionless efficiency parameter $\chi$.
Combination with $\delta z_\ell^2/t_\ell^{\mathrm{NL}}=\mathrm{const.}$ yields 
$t_\ell^{\mathrm{NL}}\sim \ell^{(1+\chi)2/3}$ compared to the standard-phenomenology result $\sim \ell^{2/g}$. 
A value of $\chi=0$ ($g=3$) yields the isotropic K41 case while 
$\chi>0$ ($g<3$) corresponds to cascade depletion and $\chi<0$ ($g>3$) to cascade enhancement.
In this picture the cascade efficiency is controlled by the factor $(\ell/\ell_0)^\chi$ in $t_\ell^{\mathrm{NL}}$, 
lumping together deviations of the physical transfer-process from the K41 picture \textit{and} differences 
in the amount of cascading energy compared to the isotropic case.
For example, the model indicates a growing field-perpendicular cascade with increasing $B_0$ though scalings suggest a transition
from K41 to the less efficient IK cascade mechanism. This efficiency reduction is, however, over-compensated by the increase of energy cascading 
field-perpendicularly compared to the isotropic situation.  
The model reproduces the numerical data very well and in agreement with the physical interpretation suggested above. 
With increasing $B_0$ a larger fraction of energy compared to the isotropic case ($B_0=0$) is spectrally transferred 
perpendicular to the magnetic field while the cascade becomes successively damped in the parallel-field direction. 

In summary, we have analyzed DNS of decaying and forced MHD turbulence without and with varying mean magnetic field 
using higher-order statistics. Globally isotropic turbulence exhibits statistical anisotropy, attributed to 
the influence of the local magnetic field on the nonlinear energy cascade. An increasing 
mean magnetic-field $B_0$ damps the parallel-field dynamics while in the perpendicular 
direction a gradual transition towards two-dimensional MHD turbulence is observed with perpendicular energy-spectra  
showing IK scaling. A modified Log-Poisson intermittency model reproduces the statistical anisotropy by phenomenological tuning of the 
respective energy cascades. 
The anisotropic approach of Goldreich and Sridhar, therefore, seems to be a promising concept though the proposed 
realizations for weak, `intermediate' and `strong' turbulence remain questionable.    
%
%%\bibliographystyle{/home/Wolf/.TeX/revtex/prsty}
%%\bibliography{/home/Wolf/Texte/Bibliographien/Turbulence}
\newcommand{\nop}[1]{}

%%%%%%%%%%%%%%%%%%%%%%%%%%%%%%%%%%%%%%%%%%%%%%%%%%%%%%%%%%%%%%%
%\newpage
\listoffigures
\end{document}